\documentclass[aps,prl,twocolumn,lengthcheck,showpacs,floatfix]{revtex4}
\usepackage{amsmath}
\usepackage{amsfonts}
\usepackage{amssymb}
\usepackage{times}
\usepackage{array}
\usepackage{graphicx}
\usepackage{ifpdf}
\usepackage[bookmarks, bookmarksnumbered,  
 pdfpagelayout=OneColumn, pdfnewwindow=true,
 pdfstartview=XYZ, plainpages=false, pdfpagelabels,
 pdfauthor={Julian A. King, John K. Webb, Michael T. Murphy and Robert F. Carswell}, pdftex,
 pdftitle={Stringent null constraint on cosmological evolution of the proton-to-electron mass ratio}]{hyperref}

\begin{document}
\title{Stringent null constraint on cosmological evolution of the proton-to-electron mass ratio}
\author{Julian~A.~King}
\author{John~K.~Webb}
\affiliation{School of Physics, University of New South Wales, Sydney, NSW, 2052, Australia}
\author{Michael~T.~Murphy}
\affiliation{Centre for Astrophysics and Supercomputing, Swinburne University of Technology, Victoria, 3122, Australia}
\author{Robert~F.~Carswell}
\affiliation{Institute of Astronomy, University of Cambridge CB3 0HA, UK}

\date{\today}

\begin{abstract}
We present a strong constraint on variation of the proton-to-electron mass ratio, $\mu$, over cosmological time scales using molecular hydrogen transitions in optical quasar spectra. Using high quality spectra of quasars Q0405$-$443, Q0347$-$383 and Q0528$-$250, variation in $\mu$ relative to the present day value is limited to $\Delta\mu/\mu=(2.6 \pm 3.0) \times 10^{-6}$. We reduce systematic errors compared to previous works by substantially improving the spectral wavelength calibration method and by fitting absorption profiles to the forest of hydrogen Lyman $\alpha$ transitions surrounding each H$_2$ transition. Our results are consistent with no variation, and inconsistent with a previous $\approx4\sigma$ detection of $\mu$ variation involving Q0405$-$443 and Q0347$-$383. If the results of this work and those suggesting that $\alpha$ may be varying are both correct, then this would tend to disfavour certain grand unification models.\end{abstract}


\maketitle
Searches have been undertaken in recent years for cosmological variations in fundamental, dimensionless constants. These searches are motivated by predictions of Kaluza--Klein theory, string theory and other grand unification theories that the so-called ``fundamental constants'' may evolve over cosmological time scales. Although much of the focus has been on $\alpha$, the fine structure constant, others have examined the proton-to-electron mass ratio, $\mu \equiv m_p/m_e$. The quantum chromodynamical scale, to which $\mu$ is sensitive, may vary faster than the quantum electrodynamical scale, hence $\mu$ may vary more than $\alpha$ \cite{Flambaum:04}. The wavelengths of the Lyman and Werner transitions of the H$_2$ molecule are sensitive to $\mu$, and examination of H$_2$ absorption systems in quasar spectra allows one to search for any such variation, as was first noted by \cite{Thompson:1975}.

Attempts from 1995 to 2004 to detect a variation in $\mu$ yielded results statistically consistent with no change \cite{Varshalovich:1993,Cowie:Songaila:1995,Varshalovich:1995,Potekhin:98,Ivanchik:02-2,Ivanchik:03-1,Levshakov:03-1,UbachsReinhold:04,Ivanchik:05}. These searches (with the exception of \citet{Ivanchik:05}) were impaired by insufficiently accurate laboratory measurements of the H$_2$ wavelengths, as well as lower quality quasar spectra. Recent laboratory XUV laser measurements \cite{CanJChem_H2,Hollenstein:06-1} have yielded substantial improvements in H$_2$ wavelength accuracy.

Using these newly available wavelengths, \citeauthor{Reinhold:06-1} (\citeyear{Reinhold:06-1}) \cite{Reinhold:06-1} reanalysed the observed H$_2$ wavelengths derived by \citet{Ivanchik:05} from Very Large Telecope (VLT) spectra of absorbers associated with Q0405$-$443 (at redshift $z$ $\approx2.595$) and Q0347$-$383 (at $z\approx3.025$), finding a change in $\mu$ of $\Delta\mu/\mu=(2.4\pm0.6)\times 10^{-5}$, where $\Delta\mu/\mu \equiv (\mu_z-\mu_0)/\mu_0$, $\mu_z$ is the measured value of $\mu$ at redshift $z$, and $\mu_0$ is the present day laboratory value. However, it has since been demonstrated \cite{Murphy:Tzanarvaris:07-1} that the techniques used to calibrate the wavelength scale of UVES (Ultraviolet and Visual Echelle Spectrograph, on the VLT) produce both long and short range calibration errors \cite{Murphy:Tzanarvaris:07-1}. These calibration errors directly impact the calculation of $\Delta\mu/\mu$. It is therefore important to re-analyse these spectra using the improved wavelength calibration techniques of \cite{Murphy:Tzanarvaris:07-1}, and we do so here. We also analyse an absorber towards Q0528$-$250 (at $z\approx 2.811$), which provides a new, strong constraint on $\Delta\mu/\mu$. We use the Voigt profile fitting program VPFIT to analyse our spectra.

For a given H$_2$ transition observed in an absorbing cloud at redshift $z_\text{abs}$, the first-order shift in the wavelength $\lambda_i$ compared to the laboratory wavelength $\lambda_0$ is given by \begin{equation}
\lambda_i=\lambda_0(1+z_{\text{abs}})(1+K_i \Delta\mu/\mu) \end{equation} where $K_i$ is the sensitivity coefficient associated with each transition, given by $K_i= (d \ln \lambda_i)/(d \ln \mu)$. $z_\text{abs}$ is the redshift of the transitions measured provided that $\Delta\mu/\mu=0$. If $\Delta\mu/\mu \neq 0$, $z_\text{abs}$ corresponds to the redshift, determined from the ensemble of available transitions, of a transitions with $K_i=0$. Previous works have calculated $K_i$ within a semi-empirical framework \cite{Reinhold:06-1,Ubachs:2007-01}; \citet{Reinhold:06-1} recently produced $K_i$ coefficients of improved accuracy by including effects beyond the Born-Oppenheimer Approximation. We use the $K_i$ coefficients calculated by \citet{Reinhold:06-1} and \citet{Ubachs:2007-01}. 

For a series of H$_2$ transitions, the best-fit value of $\Delta\mu/\mu$ may be determined in one of two ways. In the first, the ``reduced redshift method'' (``RRM''), one calculates an observed redshift $z_i$ for each transition, and then defines a reduced redshift \begin{equation}
\zeta_i=\frac{z_i-z_\text{abs}}{1+z_\text{abs}} = \frac{\Delta\mu}{\mu}K_i.
\end{equation} $\Delta\mu/\mu$ can then be determined by a linear fit to the observed $\zeta_i$ vs $K_i$ distribution.

The second method (``VPFIT method'') involves fitting all available transitions simultaneously and solving for a single redshift for each identifiable absorbing H$_2$ component in the system. We refer to each fitted redshift as a velocity component because of the close proximity of these components in velocity space. $\Delta\mu/\mu$ is estimated by perturbing the laboratory H$_2$ wavelengths as $\lambda_0 \rightarrow \lambda_0(1+K_i(\Delta\mu/\mu))$ and minimising $\chi^2$ for the spectral data fitted. The value of $\Delta\mu/\mu$ at the minimum $\chi^2$ is the best-fit value.

Although the VPFIT method has previously \cite{Potekhin:98} been used to construct $\chi^2$ vs $\Delta\mu/\mu$ curves, from which the best value of $\Delta\mu/\mu$ can be estimated, we instead include $\Delta\mu/\mu$ as a free parameter in the fit (within VPFIT), to be solved for concomitantly with the other line parameters; this yields a substantial improvement in computational speed and robustness.

The RRM has been used in most previous measurements and was the method used by \cite{Reinhold:06-1}. This method is appealing because the required numerical methods are relatively simple. However, the VPFIT method is preferable in that fewer parameters are required to fit the data. In particular, the VPFIT method has $n_v(n_t-1)$ fewer free parameters, where $n_v$ is the number of velocity components and $n_t$ is the number of transitions used. For Q0405$-$443 the VPFIT method yields 51 fewer parameters, for Q0347$-$373 it yields 67 fewer parameters, and for Q0528$-$250 it yields 252 fewer parameters.

It should be noted that the VPFIT method also improves the stability of the fitting process. In systems with multiple velocity components, particular transitions may have very poorly constrained line parameters, despite the fact that the best-fit line parameters may be well constrained over many transitions. In the RRM, this can cause the fitting algorithm to reject certain velocity components in some transitions, rendering those transitions unsuitable for inclusion in the fit. Using the VPFIT method, the reduction in the number of free parameters as a result of requiring each velocity component to occur at a single redshift helps to stabilise the fit, allowing for the inclusion of a greater number of transitions.

Each of the molecular hydrogen transitions involved falls within the Lyman $\alpha$ forest, a dense series of absorption lines blueward of the hydrogen Lyman $\alpha$ emission line of the quasar. These transitions substantially complicate the analysis; the narrow molecular hydrogen absorption lines are often situated deep within much broader, and usually complex, Lyman $\alpha$ lines. These contaminating atomic Lyman $\alpha$ transitions are insensitive to a change in $\mu$. In contrast to previous works, we fit absorption profiles to all of the Lyman $\alpha$ transitions in the vicinity of each H$_2$ transition. This allows the inclusion of a greater number of H$_2$ transitions which would otherwise be excluded from the fit.

We have reduced the total number of free parameters in the fit by tying together physically linked parameters. In particular, we tie the Doppler (linewidth) parameters together for H$_2$ transitions with the same rotational quantum number $J$ of the initial state. For the transitions we have analysed, $J \in [0,4]$. Each error estimate is multipled by $\sqrt{\chi^2_\nu}$ (where $\chi^2_\nu$ is the $\chi^2$ per degree of freedom for the whole fit), to account for a non-ideal fit.

Our inclusion of the Lyman $\alpha$ forest within the fitting process increases the number of free parameters in the fit substantially (to over 1000 in each quasar spectrum). With such a large parameter space, convergence of the optimization algorithm must be checked. \cite{Murphy:07-2,Murphy:08} demonstrated that the results of \cite{Srianand:04,Chand:2004,Levshakov:05,Levshakov:06} (in relation to $\alpha$) were flawed in this respect. Our algorithm demonstrates the proper convergence, in that $\Delta\mu/\mu$ vs $\chi^2$ curves possess the correct parabolic shape, with derived $1\sigma$ error bounds on $\Delta\mu/\mu$ that agree with those produced by VPFIT. 

The system toward Q0347$-$373 contains a single H$_2$ velocity component. The system towards Q0405$-$443 has a second velocity component, separated by $\approx 13\ \text{kms}^{-1}$ in velocity space \cite{Reinhold:06-1}. However, many of its transitions are weak or are heavily blended, and so we have not utilised the second component here.

The H$_2$ system towards Q0528$-$250 is more complicated. Previous attempts to examine this system have yielded varied and comparatively poor results \cite{Potekhin:98,UbachsReinhold:04} because the spectra used had substantially lower signal-to-noise ratio than those currently available from VLT/UVES.  \citet{Ledoux:03} report the detection of multiple velocity components in the Q0528$-$250 system (that is, multiple systems separated in velocity space), and \citet{Srianand:2005} model the absorber with 2 components.

We have tried modeling the absorber towards Q0528$-$250 with 2, 3 and 4 velocity components. Two velocity components are plainly obvious as a substantial asymmetry in every line. Using the F test, the probability that the reduction in $\chi^2$ from using three components instead of two is due to chance is $p=4\times10^{-18}$. That is, a three component model is very strongly preferred to a two component model.

Comparing a model with four velocity components to a model with three gives $p=1.8 \times 10^{-8}$. That is, the four component model is preferred over both the two and three component models. The use of a five component model produces a fit that is highly unstable numerically, and so we use the four component model as the fiducial model.

In modeling the multiple velocity components, we require that the ratio of the column densities between each velocity component is the same for transitions with the same quantum number $J$. Certain line parameters for the four-parameter fit (with appropriate redshifts tied) are given in table \ref{tab1}.

\begin{table}[h]
\begin{tabular}{c c c} 
\hline Component & Relative column density (cm$^{-2}$) & Redshift \\ \hline 
1 & $0.00 $ & $2.8110036(24)$ \\ 
2 & $-0.10 \pm 0.03 $ & $2.8111229(15)$ \\ 
3 & $-0.60 \pm 0.10$ & $2.8109334(37)$ \\ 
4 & $-1.89 \pm 0.76$ & $2.8112139(91)$ \\ \hline 
\end{tabular}
\caption{\label{tab1}Relative column densities and redshifts for the best four$-$component fit to Q0528$-$250, with $1\sigma$ uncertainties, derived from the $J=1$ set of lines. The relative column density is the difference between the logarithm of the column density for each component and the logarithm of the column density for the strongest component. We use the $J=1$ set of lines because they are the largest fraction of the data set.}
\end{table}

Our results are set out in tables \ref{tab2} and \ref{tab3}.
%Results

\begin{table}[h]
\begin{tabular}{c c c c c} 
\hline Quasar spectrum & $\Delta\mu/\mu$ --- VPFIT & $\chi^2_\nu$ & $z$ & $n$ \\ \hline 
Q0405$-$443 & $(10.1 \pm 6.2) \times 10^{-6}$ & $1.42$ & 2.595 & 52\\ 
Q0347$-$373 & $(8.2 \pm 7.4) \times 10^{-6}$ & $1.28$ & 3.025 & 68\\ 
Q0528$-$250 & $(-1.4 \pm 3.9) \times 10^{-6}$  & $1.22$ & 2.811 & 64 \\  
Weighted mean & $(2.6 \pm 3.0) \times 10^{-6}$  & n/a & $2.81$ & n/a\\  \hline
\end{tabular}
\caption{\label{tab2}Values of $\Delta\mu/\mu$ obtained using the VPFIT method, derived from each of the quasar spectra. $n$ is the number of transitions (per velocity component, for Q0528$-$250). The weighted mean given here is our preferred result.}
\end{table} 

\begin{table}[h]
\begin{tabular}{c c c c c} 
\hline Quasar spectrum & $\Delta\mu/\mu$ --- RRM & $\chi^2_\nu$ & $z$ & $n$\\ \hline 
Q0405$-$443 & $(10.9 \pm 7.1) \times 10^{-6}$ & $1.01$ & 2.595 & 52\\ 
Q0347$-$373 & $(6.4 \pm 10.3) \times 10^{-6}$ & $1.13$ & 3.025 & 68 \\ 
Q0405 + Q0347 & $(8.5 \pm 5.7) \times 10^{-6}$ & $1.06$ & 2.810 & 120\\
Q0528$-$250 & n/a  & n/a & n/a & 64 \\  \hline
\end{tabular}
\caption{\label{tab3}Values of $\Delta\mu/\mu$ using the RRM, derived from each of the quasar spectra. $n$ is the number of transitions.}
\end{table} 

We re-sampled the $\zeta_i$ vs $K_i$ graph with the bootstrap method \cite{NumericalRecipes:2002}, to check for consistency. That is, we generated $10^5$ new data sets by randomly drawing data points, with replacement, from the original data set, such that each of the new data sets has the same number of points as the original data set. We then obtained the slope of the linear fit to each of these data sets, giving $\Delta\mu/\mu$ for each set; the mean of this ensemble should be consistent with the generating data. For each absorber, we found consistency between the bootstrap method, the VPFIT method and the RRM.

The RRM is not appropriate for Q0528$-$250 because the line parameters for each of the velocity components within a given transition are strongly correlated. So, for Q0528$-$250, groups of 4 points in a $\zeta_i$ vs $K_i$ plot (the RRM method) are not independent. Thus the RRM method breaks down, as it uses linear least squares fitting, which requires independence of all the data points. This demonstrates the superiority of the VPFIT method over the RRM. These correlations are correctly incorporated into the VPFIT method, as $\Delta\mu/\mu$ is determined from $\chi^2$ on the total model.

We checked that using 4 velocity components (instead of 2 or 3) has no significant effect on the result by solving for $\Delta\mu/\mu$ (within VPFIT), with 2 and 3 velocity component models. This produces $\Delta\mu/\mu=(-0.6 \pm 3.8) \times 10^{-6}$ and $\Delta\mu/\mu=(-1.4 \pm 3.8) \times 10^{-6}$ respectively. These are similar to each other and to the result for the four velocity component model. This demonstrates the insensitivity of the result to having used the statistically preferred velocity structure.

Combining the three measurements of $\Delta\mu/\mu$ obtained within VPFIT using a weighted mean yields the value $\Delta\mu/\mu = ( 2.6 \pm 3.0) \times 10^{-6}$. This is null at a $1\sigma$ confidence level. This is our main result from a combined analysis of all three quasar absorbers.

\ifpdf
\begin{figure*}
\includegraphics[angle=-90,viewport=42 23 188 477,width=172mm]{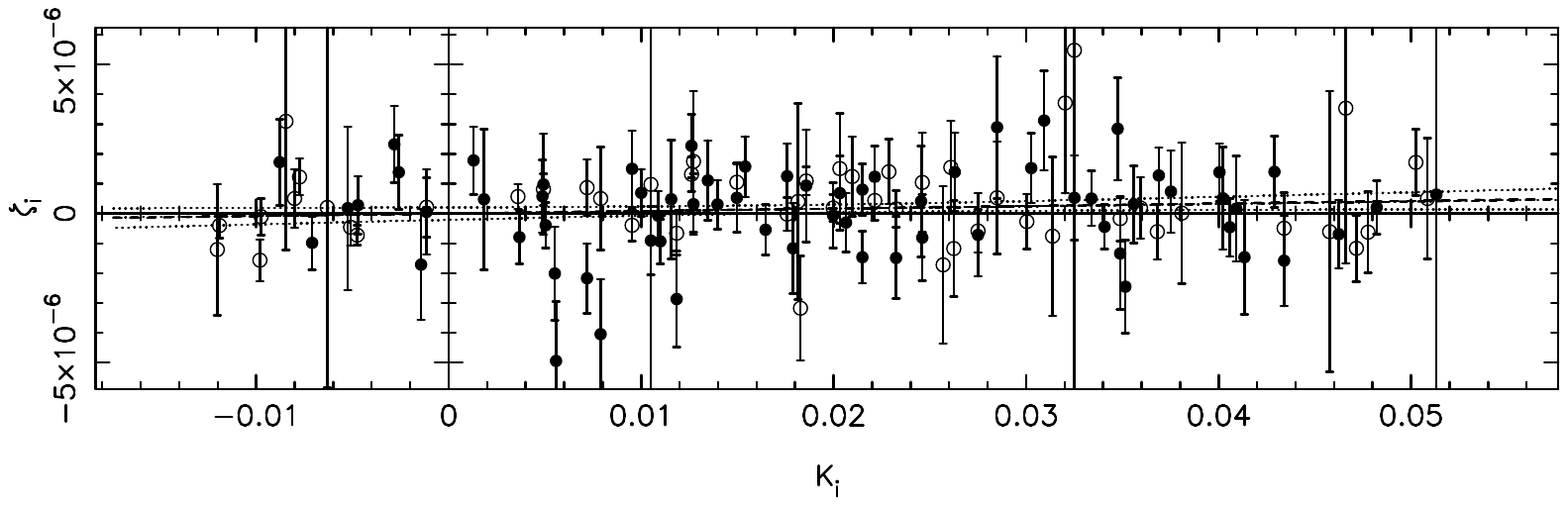}
\caption{\label{fig1}Reduced redshift plot ($\zeta_i$ vs $K_i$) for Q0405$-$443 and Q0347$-$373 with gradient $\Delta\mu/\mu=(8.5 \pm 5.7)\times 10^{-6}$ (dashed line). Q0347$-$373 is represented by closed circles, and Q0405$-$443 is represented by open circles. The unweighted fit, a dot-dashed line, is obscured by the weighted fit. The dotted lines give the $1\sigma$ confidence limit on the regression line. Note that this is not our preferred result, because it does not include Q0528$-$250 in the fit. This graph may be compared directly with fig. 2 of \cite{Reinhold:06-1}.}
\end{figure*}
\else
\begin{figure*}
\includegraphics[angle=-90,viewport=42 23 188 477,width=172mm]{graphics/Q0347+Q0405_reducedredshift_2c.eps}
\caption{\label{fig1}Reduced redshift plot ($\zeta_i$ vs $K_i$) for Q0405$-$443 and Q0347$-$373 with gradient $\Delta\mu/\mu=(8.5 \pm 5.7)\times 10^{-6}$ (dashed line). Q0347$-$373 is represented by closed circles, and Q0405$-$443 is represented by open circles. The unweighted fit, a dot-dashed line, is obscured by the weighted fit. The dotted lines give the $1\sigma$ confidence limit on the regression line. Note that this is not our preferred result, because it does not include Q0528$-$250 in the fit. This graph may be compared directly with fig. 2 of \cite{Reinhold:06-1}.}
\end{figure*}
\fi

For comparison with \citet{Reinhold:06-1}, a reduced redshift plot (Fig.~\ref{fig1}) including only Q0405$-$443 and Q0347$-$373 produces the result $\Delta\mu/\mu=(8.5 \pm 5.7)\times 10^{-6}$ (weighted fit) and $\Delta\mu/\mu=(7.9 \pm 8.1)\times 10^{-6}$ (unweighted fit). We also attempt to compare with \cite{Reinhold:06-1} by including in the fit only those transitions used in that paper. For Q0405$-$443, this removes 16 transitions and adds 3, the latter of which appear to be contaminated and which were excluded from our main analysis. This yields, from the RRM, a Q0405$-$443 result of $\Delta\mu/\mu=(10.2\pm8.9)\times 10^{-6}$. This is offset from the result of $(27.8 \pm 8.8)\times 10^{-6}$ in \cite{Reinhold:06-1}. For Q0347$-$373, we remove 35 transitions that are not used in \cite{Reinhold:06-1}, and include 4 which appear to be contaminated, to give a result of $\Delta\mu/\mu=(12.0 \pm 14.0) \times 10^{-6}$, compared with $(20.6 \pm 7.9)\times 10^{-6}$ from \cite{Reinhold:06-1}. The weighted mean of our results in this circumstance is $\Delta\mu/\mu=(10.7 \pm 7.5)\times 10^{-6}$. It is difficult to make a direct statistical comparison, due to the fact that the spectra analysed are not independent, however in both cases we see a shift of $\Delta\mu/\mu$ towards $0$. Although the inclusion of Q0528$-$250 clearly shifts the combined Q0405$-$443 + Q0347$-$373 result towards zero, our combined Q0405$-$443 + Q0347$-$373 result is null under all the circumstances considered.

\begin{figure*}[htbp!]
\includegraphics[angle=-90,viewport=52 24 275 472,width=172mm]{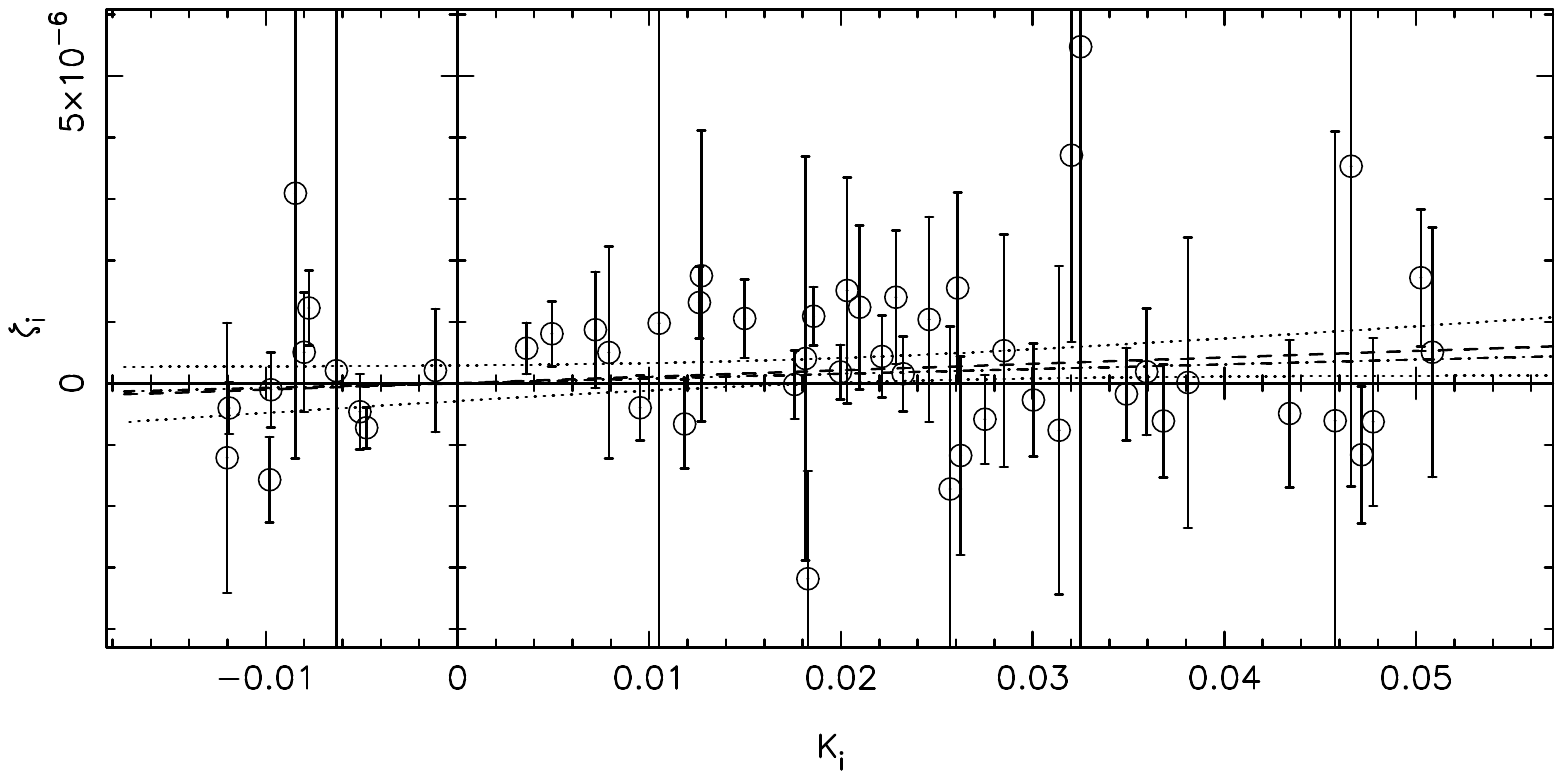}
\caption{\label{Q0405_reducedredshift}Reduced redshift ($\zeta_i$ vs $K_i$) for Q0405-443. This graph has gradient $\Delta\mu/\mu=(1.09 \pm 0.71) \times 10^{-5}$ (weighted fit). The weighted, line of best fit is given by the dashed line with the unweighted line of best given by the dot-dashed line. The dotted lines indicate the 1$\sigma$ (weighted) confidence intervals on the regression line. This figure is not included in the shorter version of this paper.}
\end{figure*}

\begin{figure*}[htbp!]
\includegraphics[angle=-90,viewport=52 24 275 472,width=172mm]{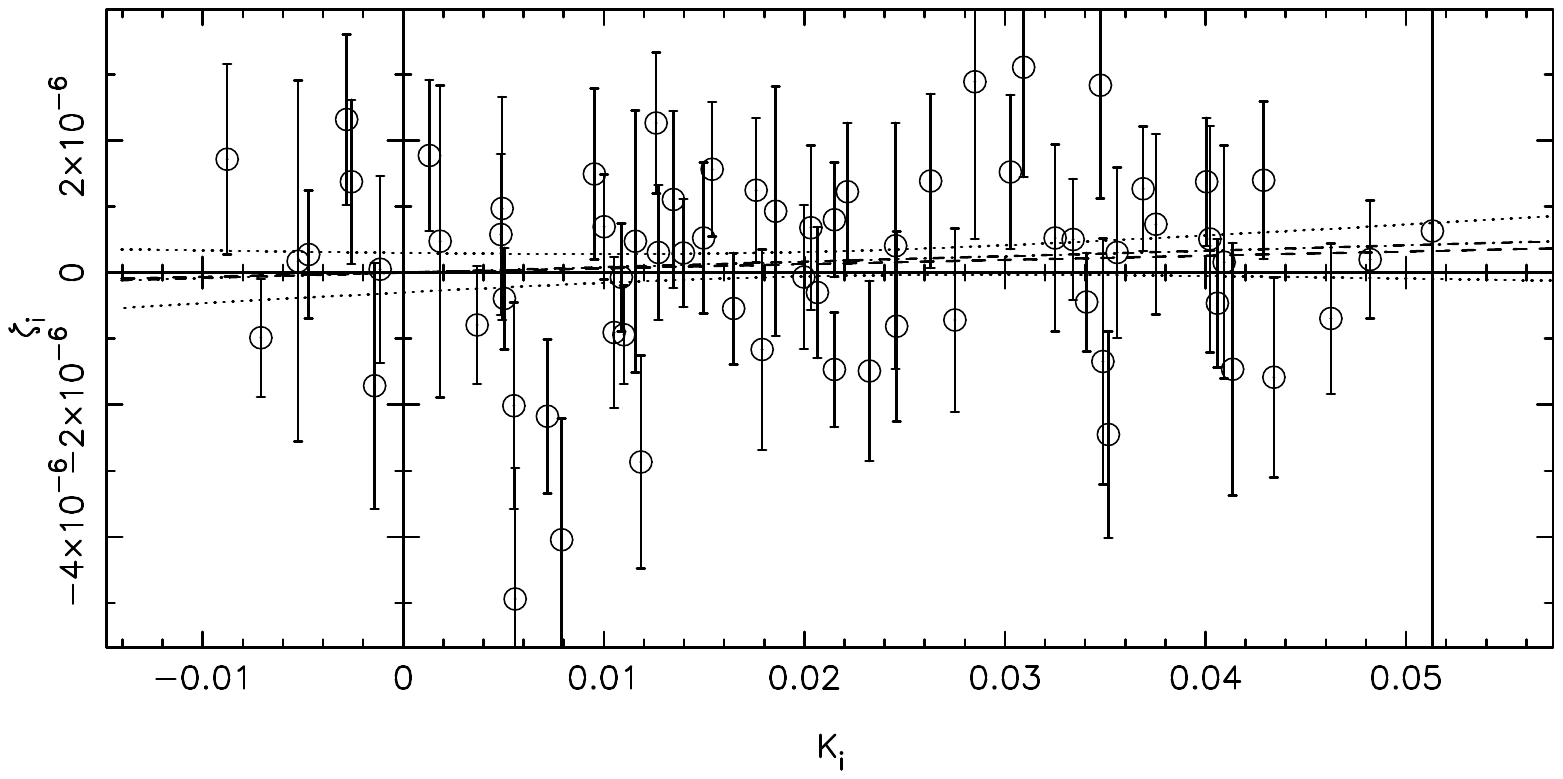}
\caption{\label{Q0347_reducedredshift}Reduced redshift ($\zeta_i$ vs $K_i$) for Q0347-373. This graph has gradient $\Delta\mu/\mu=(6.4\pm 10.3)\times 10^{-6}$ (weighted fit). The weighted, line of best fit is given by the dashed line with the unweighted line of best given by the dot-dashed line. The dotted lines indicate the 1$\sigma$ (weighted) confidence intervals on the regression line. This figure is not included in the shorter version of this paper.}
\end{figure*}

\begin{figure*}[htbp!]
\includegraphics[viewport=0 0 346 244,width=150mm]{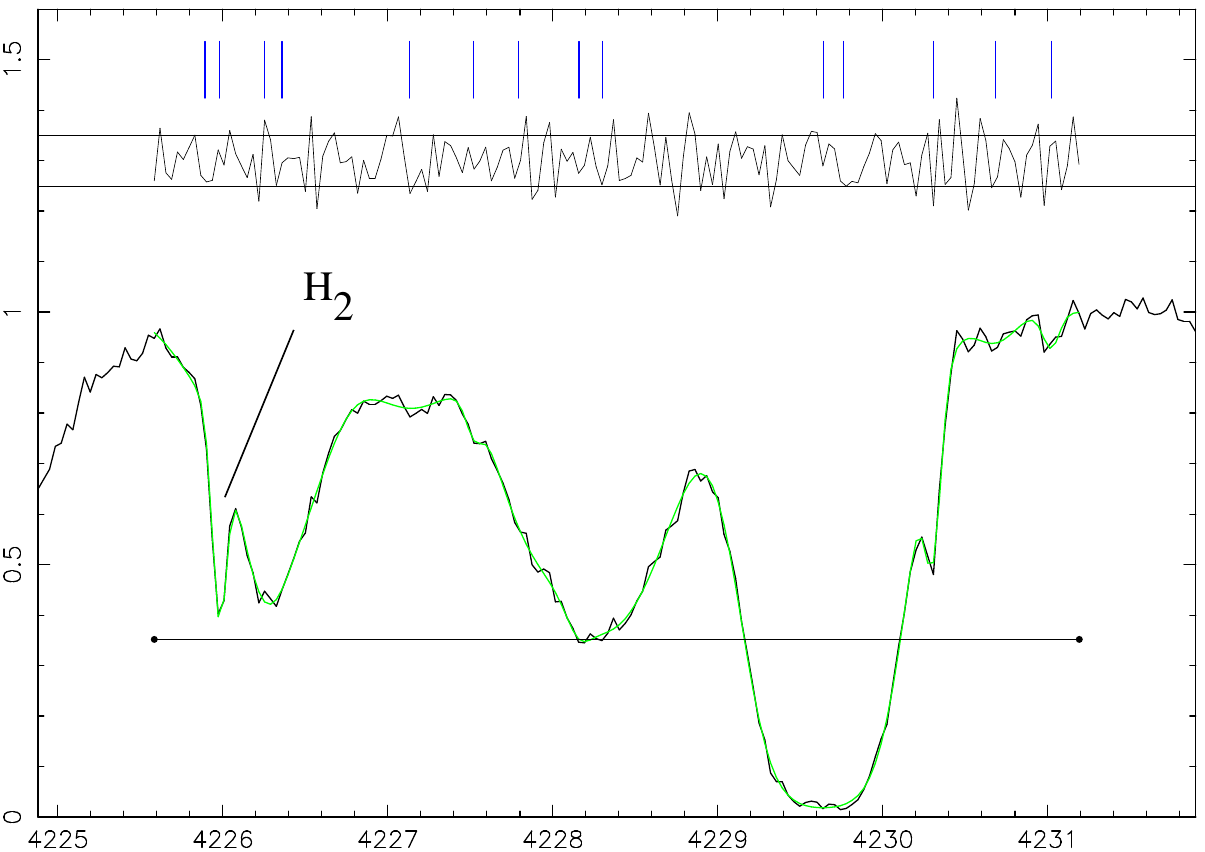}
\caption{\label{Q0347r12}Fitting region 12 of 45 for Q0347$-$373. The centre positions of transitions are indicated by the blue tick marks. The top plot indicates the residuals, normalised to $1\sigma$ (the horizontal lines). In this case, the second tick mark from the left ($\approx 4226 \text{\AA}$) is a H$_2$ transition. All lines within the fitting region except those indicated are fitted as hydrogen Lyman $\alpha$. This figure is not included in the shorter version of this paper.}
\end{figure*}

\begin{figure*}[htbp!]
\includegraphics[viewport=0 0 511 360,width=150mm]{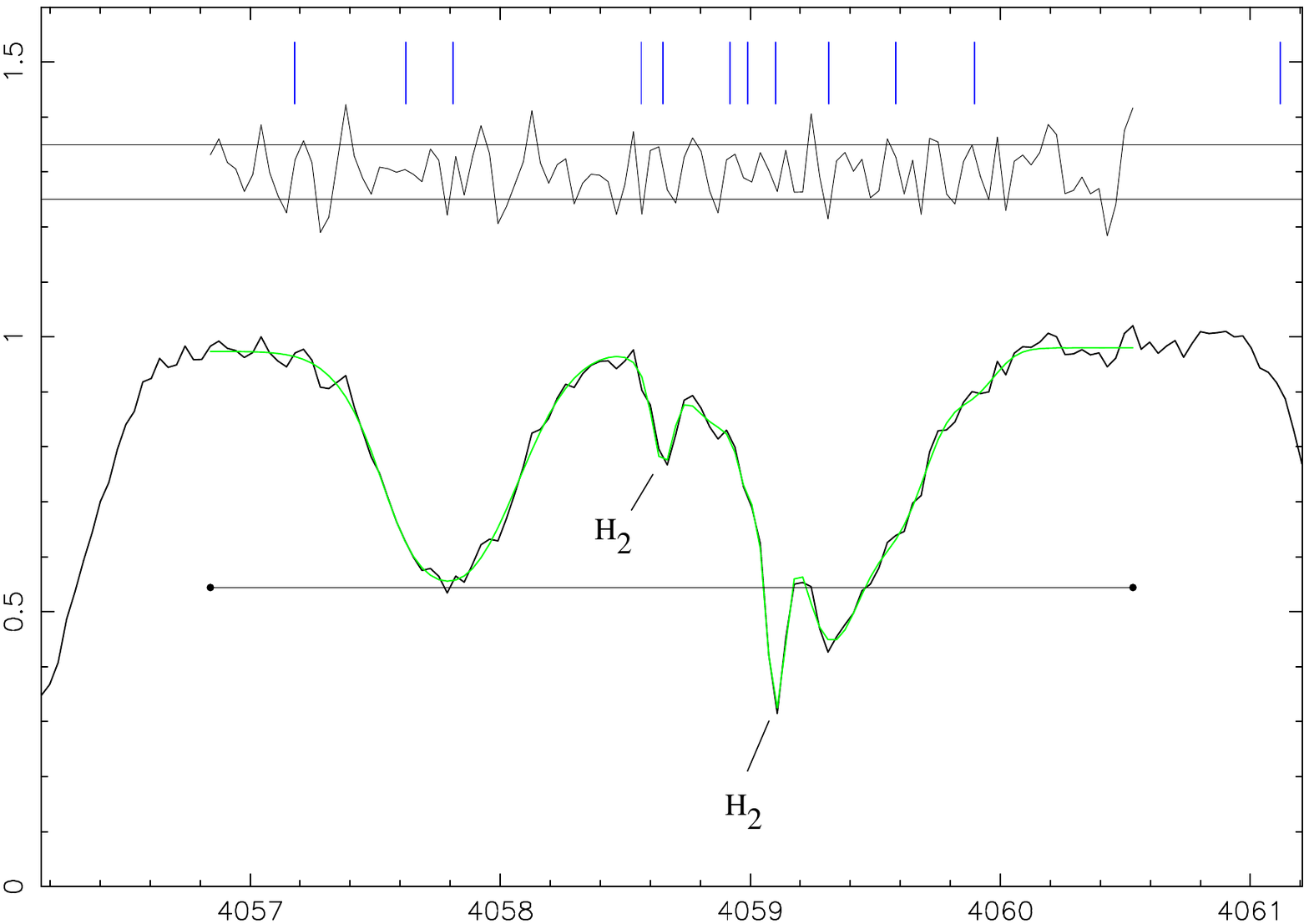}
\caption{\label{Q0347r24}Fitting region 24 of 45 for Q0347$-$373. The centre positions of transitions are indicated by the blue tick marks. The top plot indicates the residuals, normalised to $1\sigma$ (the horizontal lines). In this case, the sharp transitions located at $\approx 4058.6\text{\AA}$ and $\approx 4059.1\text{\AA}$ are H$_2$ transitions. All lines within the fitting region except those indicated are fitted as hydrogen Lyman $\alpha$. This figure is not included in the shorter version of this paper.}
\end{figure*}

\begin{figure*}[htbp!]
\includegraphics[viewport=0 0 467 329,width=150mm]{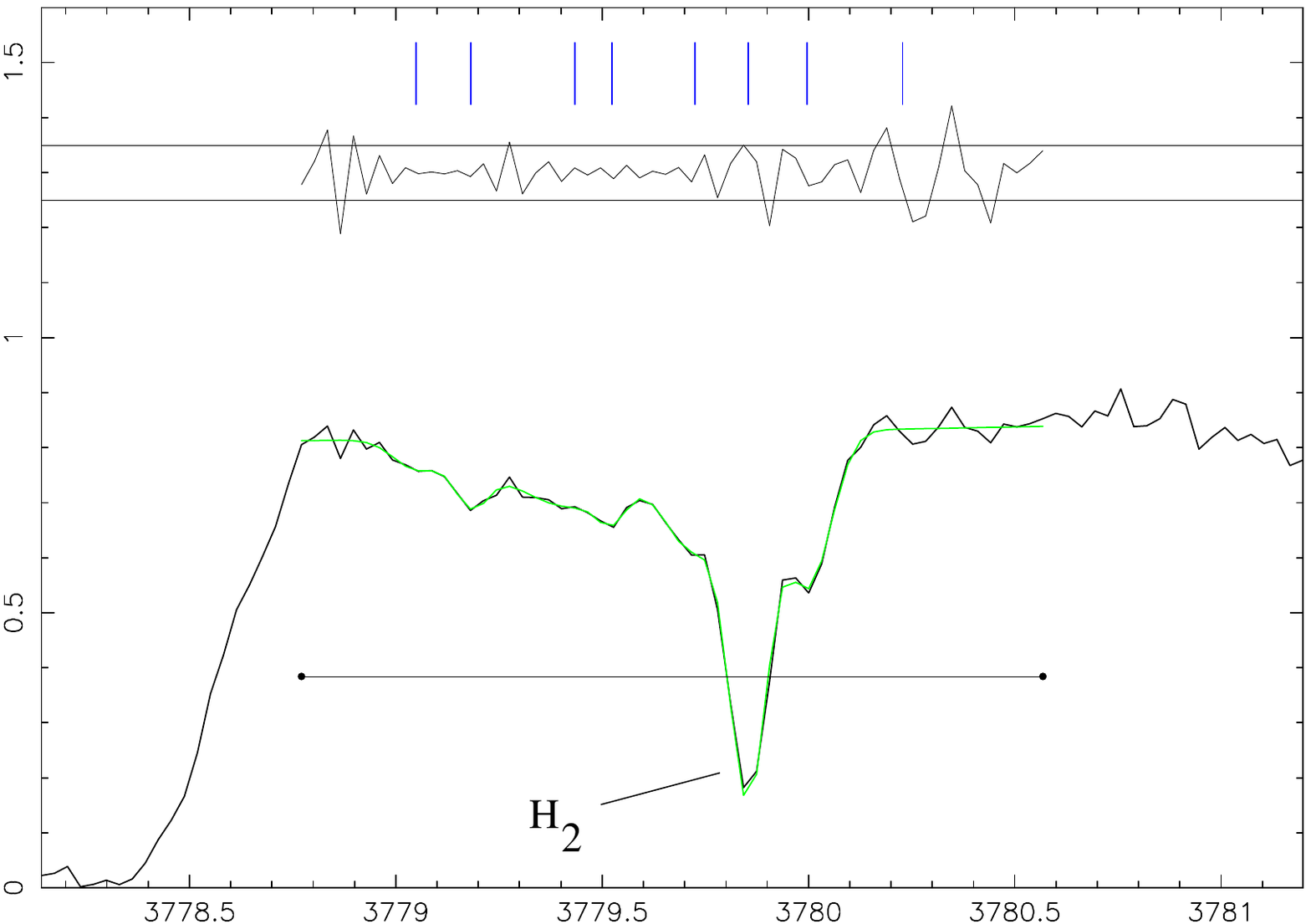}
\caption{\label{Q0405r14}Fitting region 14 of 45 for Q0405$-$443. The centre positions of transitions are indicated by the blue tick marks. The top plot indicates the residuals, normalised to $1\sigma$ (the horizontal lines). In this case, the sharp transition located at $\approx 3779.85\text{\AA}$ is a H$_2$ transition. All lines within the fitting region except those indicated are fitted as hydrogen Lyman $\alpha$. This figure is not included in the shorter version of this paper.}
\end{figure*}

\begin{figure*}[htbp!]
\includegraphics[viewport=0 0 451 318,width=150mm]{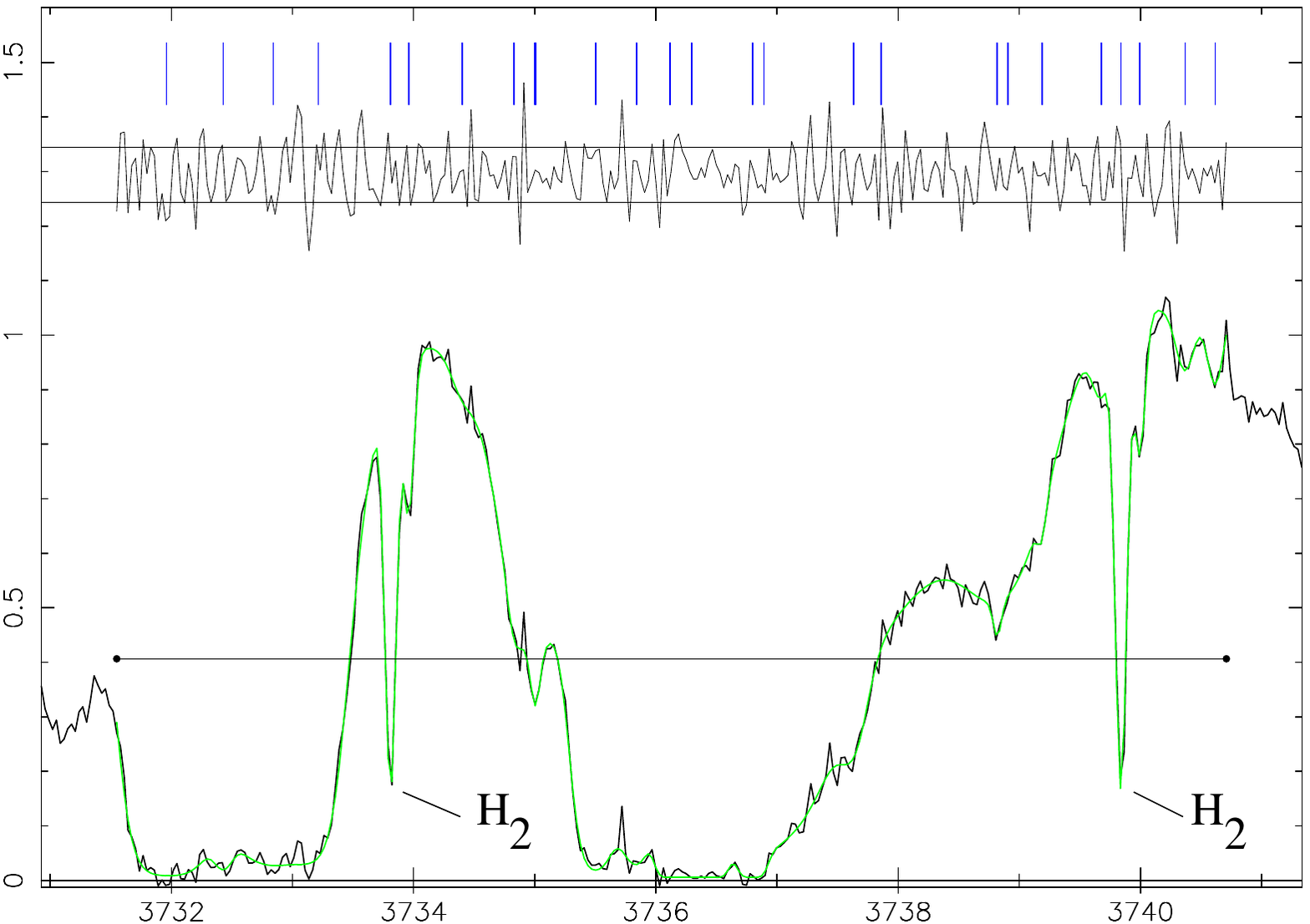}
\caption{\label{Q0405r16}Fitting region 16 of 45 for Q0405$-$443. The centre positions of transitions are indicated by the blue tick marks. The top plot indicates the residuals, normalised to $1\sigma$ (the horizontal lines). In this case, the very prominent, sharp transitions located at $\approx 3733.8\text{\AA}$ and $\approx 3739.9\text{\AA}$ are H$_2$ transitions. All lines within the fitting region except those indicated are fitted as hydrogen Lyman $\alpha$. This figure is not included in the shorter version of this paper.}
\end{figure*}

\begin{figure*}[htbp!]
\includegraphics[viewport=0 0 390 275,width=150mm]{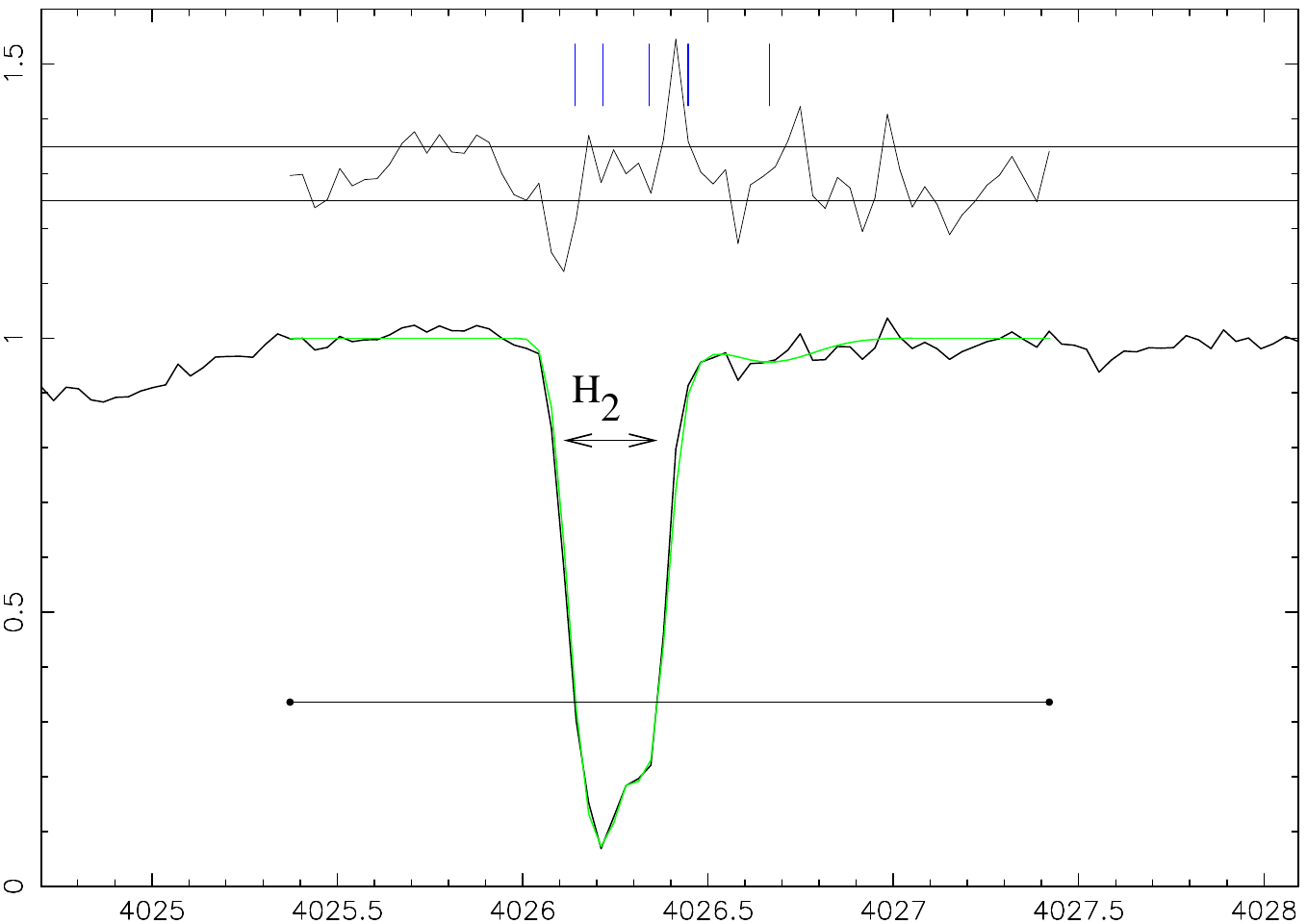}
\caption{\label{Q0528r11}Fitting region 11 of 49 for Q0528$-$250. The centre positions of transitions are indicated by the blue tick marks. The top plot indicates the residuals, normalised to $1\sigma$ (the horizontal lines). In this case, the four tick tick marks from the left indicate the four H$_2$ velocity components of Q0528$-$250, with the right-most tick mark being a Lyman $\alpha$ transition. This figure is not included in the shorter version of this paper.}
\end{figure*}

\begin{figure*}[htbp!]
\includegraphics[viewport=0 0 410 287,width=150mm]{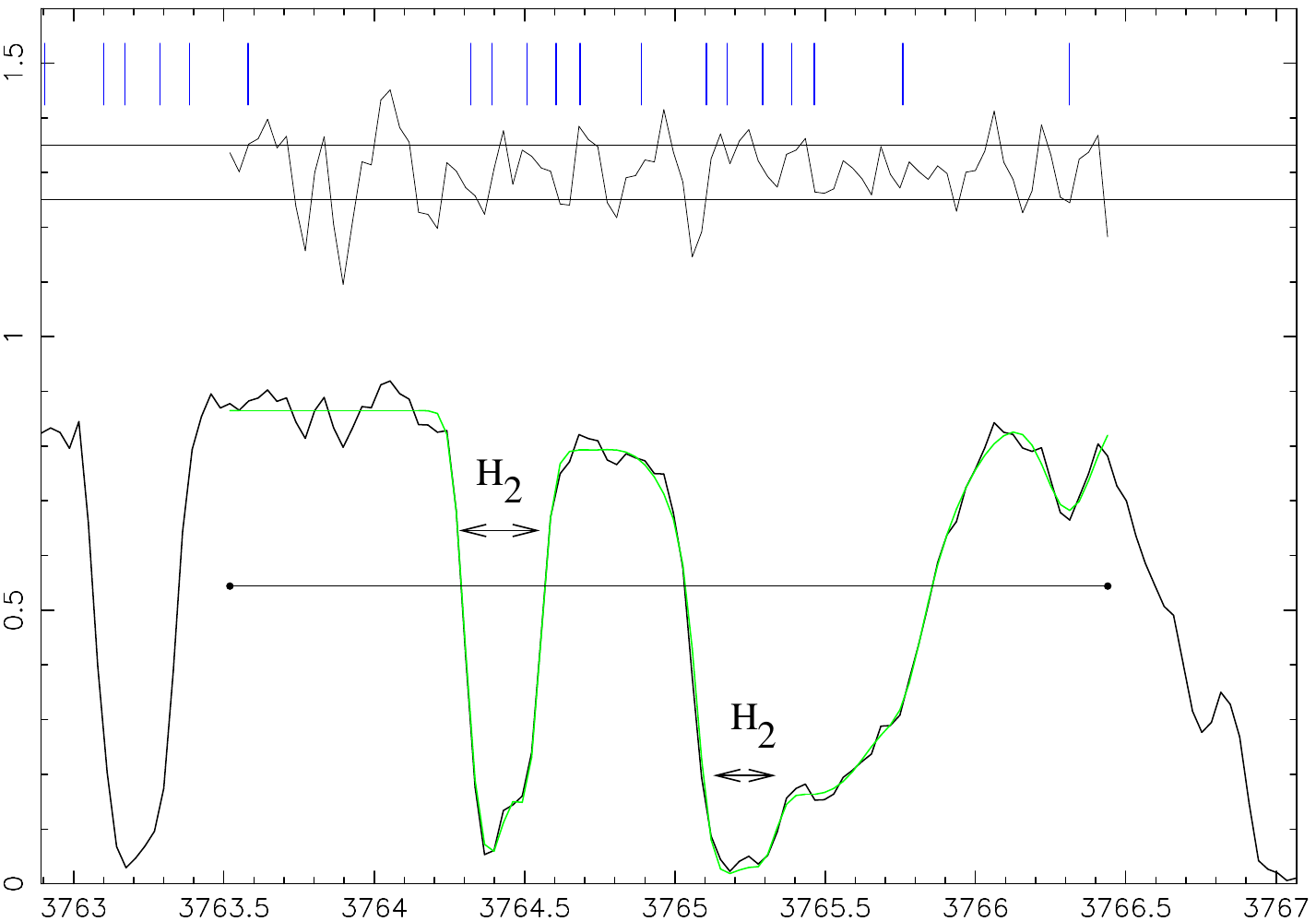}
\caption{\label{Q0528r24}Fitting region 24 of 49 for Q0528$-$250. The centre positions of transitions are indicated by the blue tick marks. The top plot indicates the residuals, normalised to $1\sigma$ (the horizontal lines). In this case, there are two sets of H$_2$ transitions, with four velocity components each, clustered about $\approx 3764.4\text{\AA}$ and $\approx 3765.2\text{\AA}$. There is an additional set of H$_2$ transitions at $\approx 3763.2\text{\AA}$, however these are not included within this fitting region, but rather in the adjacent one. All lines within the fitting region except those indicated are fitted as hydrogen Lyman $\alpha$. This figure is not included in the shorter version of this paper.}
\end{figure*}

\begin{figure*}[htbp!]
\includegraphics[viewport=19 167 545 551,width=150mm]{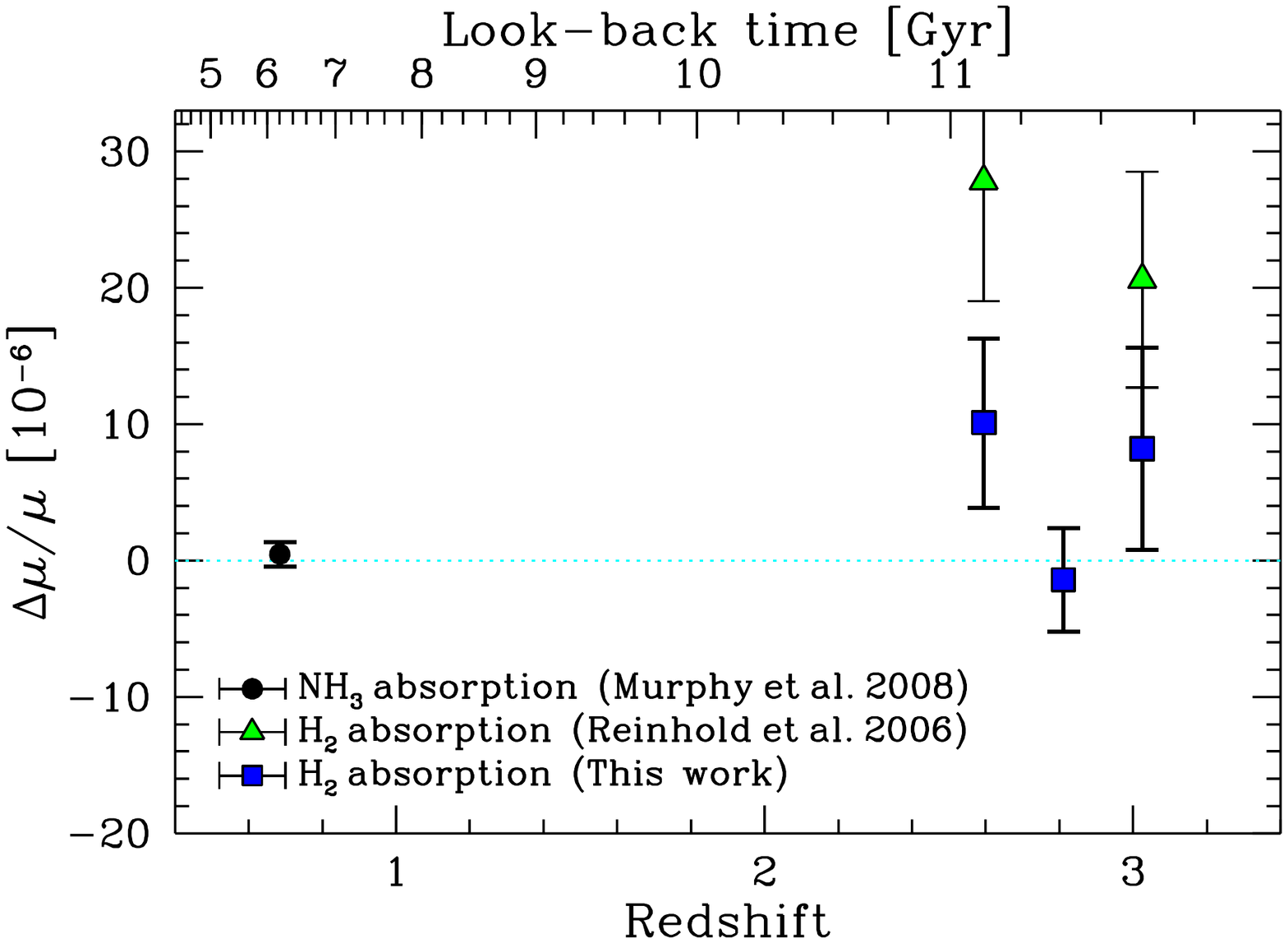}
\caption{\label{mu_vs_z_Reinhold}A comparison of the results of this work, with that of \citet{Reinhold:06-1} and \citet{Murphy:Flambaum:08}.}
\end{figure*}

%More details regarding our results can be found in \cite{KingWebb:08:arXiv:01}. In particular, the reader may locate individual copies of our $\zeta_i$ vs $K_i$ plots there, as well as some example spectra fits. \cite{KingWebb:08:arXiv:01} does not contain any information important to our result that is not included here.
Figures \ref{Q0405_reducedredshift} and \ref{Q0347_reducedredshift} show reduced redshift plots not included in the shorter, published version of this paper. Example fits also not in the shorter version of this paper can be found in figs \ref{Q0347r12}, \ref{Q0347r24}, \ref{Q0405r14}, \ref{Q0405r16}, \ref{Q0528r11}, \ref{Q0528r24} and \ref{mu_vs_z_Reinhold}.

Our final result of $\Delta\mu/\mu = ( 2.6 \pm 3.0) \times 10^{-6}$ represents a significant increase in precision over previous works (a factor of $\approx 2$). This result is entirely consistent with $\Delta\mu/\mu=0$ over cosmological timescales. It is also consistent with the recently-published work of \citet{Murphy:Flambaum:08}, who find that $\Delta\mu/\mu = (0.74 \pm 0.47_\text{stat}\pm 0.76_\text{sys})\times 10^{-6}$ using the inversion transitions of ammonia. Note, however, that the ammonia constraint is at $z=0.685$ while all our constraints are at $z>2.5$; they may not be directly compared without a theory of cosmologically evolving $\mu$. 

The unification of all interactions clearly requires that any cosmological variations in the various fundamental constants will be linked to each other. Grand Unified theories typically predict $\Delta \mu / \mu \approx R \Delta \alpha / \alpha$ \cite{Calmet:Fritsch:2001-1,Calmet:2002-1,Langacker:2002-1}, where both the sign and magnitude of $R$ are strongly model dependent. $\lvert R \rvert\approx 30-40$ emerges from many GUT models \cite{Calmet:Fritsch:2001-1,Calmet:2002-1,Langacker:2002-1}. Generally speaking, $\lvert R \rvert \gg 1$. The most reliable constraint on $\alpha$ variation at present is $\Delta\alpha/\alpha = (-5.7\pm1.0)\times 10^{-6}$ \cite{Murphy:2003-1}; the works of \cite{Srianand:04,Chand:2004,Levshakov:05,Levshakov:06} and others have been demonstrated to be unreliable \cite{Murphy:07-2,Murphy:08}. Taking both this and our new null result at face value, any variation in $\mu$ is almost 2 orders of magnitude below that expected on the basis of the $\alpha$-variation results. If both these results are both correct, those Grand Unified models which predict $\lvert R \rvert \gg 1$ are disfavoured.

%%Insert extra graphics here%%

%%Now have some selected fits%%

\begin{acknowledgments}
We would like to acknowledge V. V. Flambaum, for advice and support in various stages of this work.
\end{acknowledgments}

\bibliography{PRL_v01.bib}
\end{document}